\documentclass[%
amsmath,
amssymb,
]{revtex4-2}

\usepackage{caption}
\usepackage{subcaption}
\usepackage{graphicx}
\usepackage{dcolumn}
\usepackage{bm}
\usepackage{color, xcolor}
\usepackage{booktabs}
\usepackage{amsmath}
\usepackage{siunitx}
\usepackage{comment}
\usepackage[T1]{fontenc}
\usepackage{listings}
\usepackage{tikz}
\usepackage{booktabs} 
\usepackage{tabularx}
\usetikzlibrary{shapes.multipart, positioning}
\usepackage{makecell}
\usepackage[colorinlistoftodos]{todonotes} 
\usepackage{lipsum}
\usepackage{url}
\usepackage{hyperref}
\hypersetup{
  colorlinks=true
}

\lstdefinelanguage{yaml}{
  keywords={true,false,null},
  sensitive=false,
  comment=[l]{\#},
  morestring=[b]",
}

\begin{document}

\preprint{APS/123-QED}

\title{A flexible start-to-end simulation framework for particle accelerators based on a comprehensive lattice description}

\author{A.~D.~Brynes}\email[]{alexander.brynes@stfc.ac.uk}
\affiliation{Accelerator Science \& Technology Centre, STFC Daresbury Laboratory, Warrington, United Kingdom}
\affiliation{Cockcroft Institute, Warrington, United Kingdom}
\author{J.~K.~Jones}
\affiliation{Accelerator Science \& Technology Centre, STFC Daresbury Laboratory, Warrington, United Kingdom}
\affiliation{Cockcroft Institute, Warrington, United Kingdom}
\author{M.~King}
\affiliation{Accelerator Science \& Technology Centre, STFC Daresbury Laboratory, Warrington, United Kingdom}
\affiliation{Cockcroft Institute, Warrington, United Kingdom}
\author{M.~A.~Johnson}
\affiliation{Accelerator Science \& Technology Centre, STFC Daresbury Laboratory, Warrington, United Kingdom}
\affiliation{Cockcroft Institute, Warrington, United Kingdom}
\author{N.~Ziyan}
\affiliation{Accelerator Science \& Technology Centre, STFC Daresbury Laboratory, Warrington, United Kingdom}
\affiliation{Cockcroft Institute, Warrington, United Kingdom}

\date{\today}

\begin{abstract}
Standardization of data formats in a scientific discipline brings a range of benefits to researchers, as it enables the sharing of workflows and solutions to common problems, provides the foundation for generically useful tools that can be applied across the field, and gives a basis for cross-checking and validation that can be understood by all. Owing to the wide range of possible modes of description of particle accelerator lattices, a standard solution to this problem has not yet been developed for the field, although efforts are underway across the community. This article presents a schema for a comprehensive and generic format for describing particle accelerator lattices, encompassing physical element information, simulation code-specific parameters, control system variables, electrical and magnetic data, and other parameters, for each element. A translation layer is also provided in order to export this lattice into formats suitable for a variety of standard accelerator simulation codes. Based on this format, a framework has been developed for generating, tracking and analyzing beams through the lattice, providing a seamless transfer between simulation codes and the basis for a fully generic start-to-end simulation framework. 
\end{abstract}

\maketitle

\section{Introduction}


The components that define a particle accelerator lattice can be described in a multitude of ways, depending on the perspective of the user and the level of description required. Accelerator physicists typically focus on the effect of these components on the properties of beams propagating through a physical lattice; machine operators need to understand the methods for interacting with these components to run the accelerator via a control system; mechanical engineers must know the physical locations of the lattice elements, including any errors; magnet and RF engineers have in-depth knowledge of specific components. A common issue that arises during various stages of the life-cycle of a particle accelerator is defining the ground truth of the lattice, with different experts referring to the same set of elements in various ways. Without a base-level description of the lattice, these different descriptions may diverge: if a particular component is found to be misaligned in the physical lattice, for example, how does this information cascade so that the physicist can adjust their model? How do readings from the control system connect to the beam physics?

The development of a start-to-end simulation framework likewise depends on a standard description of a particle accelerator lattice. Many facilities make use of a lattice model from a particular simulation code that can describe the lattice components at a reasonably comprehensive level, even if that code is not suitable for simulating all of the appropriate beam physics modalities. This model is then used to derive lattice files for other simulation codes, often manually. Some important issues arise from using a lattice format derived from an accelerator code as the ground truth. Firstly, there are very few (if any) codes that provide a sufficiently comprehensive description of all accelerator lattice components, particularly if one considers elements such as photon transport beamlines or plasma cells as part of the lattice. Secondly, there is often no clearly defined route to cascading external information, such as control system parameters and mechanical measurements, down to this lattice. Similar issues arise when describing the lattice from another perspective, such as in terms of mechanical or control system parameters. 

There is, therefore, a clear need for an unambiguous and comprehensive definition of a particle accelerator lattice and its constituent elements. Standardization of particle accelerator lattice formats has been attempted in the past, with the Accelerator Markup Language (AML) and the Unified Accelerator Libraries (UAL) projects being among the most notable examples \cite{EPAC2006.WEPCH150,Part.Accel.55.313,AIPConf.391.337,PAC2003.TOPB004,IPAC2011.WEPB121}. While these formats were relatively comprehensive, they were not widely adopted beyond a handful of facilities, in part owing to difficulties in sharing code at the time, and a lack of standardised practices and formats. The attempt at lattice standardisation has seen a recent resurgence, with the Particle Accelerator Lattice Standard (PALS) collaboration beginning to formulate a universal standard \cite{PALS,NAPAC2025.TUP004}. This article presents a schema -- dubbed LAURA (Lattice Architecture for a Unified Representation of Accelerators) \cite{LAURA} -- which provides a comprehensive description of accelerator elements and lattices, aiming to extend beyond previous implementations. This standard is outlined in Sec.\,\ref{sec:laura}. While the ultimate standard format for accelerator lattices should be agnostic to the programming language and file format used, this initial attempt at standardization is based on Python objects and YAML or JSON files, with each file or object describing, in as comprehensive a manner as possible, individual elements and machine sections and lattices. Once this standard format is established, the schema can be trivially extended towards interpretations in other programming languages and storage in various file formats. 

This description of an accelerator can be used to produce lattice files for a range of simulation codes (see Sec.\,\ref{sec:translation}), and to provide the foundation for a start-to-end simulation framework. A great number of simulation codes are available for simulating particle accelerators (see Ref.\,\cite{SimulationCodes} for a list of examples), all of which have their own benefits and drawbacks, depending on the particular applications for which they are intended. General features which are common to most codes include (but are not limited to) the tracking of the 6D phase space (three co-ordinates each in position and momentum) of a particle, or a number of macroparticles, through standard accelerator elements modelled as electric and magnetic fields. Depending on the parameters of the particle bunch and the accelerator, collective effects such as space-charge (at low values of the relativistic Lorentz factor) and synchrotron radiation (in dispersive regions) can become important, and certain codes are able to simulate these interactions with varying degrees of accuracy. 

In many cases, therefore, start-to-end simulations of particle accelerators are not performed with a single code, and each of these codes typically requires input, lattice, and beam distribution files to be prepared in idiosyncratic formats. Tools which are able to handle the preparation of these files, and to pass beam distributions between them, are therefore incredibly useful for those who study these machines, as it alleviates the burden of preparing such files by hand, or with custom scripts. Two prominent recent examples of start-to-end simulation toolkits include the Lightsource Unified Modelling Environment (LUME) \cite{IPAC2021.THPAB217,LUME}, and the Adaptable Beginning-to-End Linac Framework (ABEL) \cite{IPAC2025.TUPS012,ABEL}. LUME provides wrappers for various simulation codes, covering photoinjectors, linacs and free-electron lasers (FELs), in addition to modules for photon transport and control system interactions, and it has been adopted at numerous facilities. LUME depends on template input files for some of these codes, however, leaving potentially unanswered the question of how to cascade information through to all simulation codes from a ground source of truth about the lattice. ABEL provides a range of simulation modalities, covering linacs, plasma acceleration, and beam-beam interactions, employing various levels of simulation complexity. In its current state, ABEL uses `reduced models' for simulating certain parts of an accelerator, and is well-suited for fast optimizations of certain machine sections, while also providing the functionality for more detailed investigations of plasma-based acceleration. As with LUME, ABEL does not solve the question of information cascade, and requires specific element definitions. Neither frmework solves the ground-truth issue in an easil;y extensible way.

Based on the comprehensive description of an accelerator lattice provided by LAURA, a start-to-end simulation framework has been developed to attempt to solve these problems: SIMBA -- Simulations for Integrated Modeling of Beams in Accelerators \cite{SIMBA}; see Sec.\,\ref{sec:simba}. This is a Python-based package that can read in the full set of lattice element files from an instance of LAURA and construct input, lattice and beam distribution files for a variety of simulation codes dynamically, including ASTRA \cite{ASTRA}, General Particle Tracer (GPT) \cite{GPT}, ELEGANT \cite{Elegant,APS-LS-287}, OCELOT \cite{NIMA.768.1.151,OCELOT}, Cheetah \cite{PhysRevAccelBeams.27.054601,Cheetah} CSRTrack \cite{FEL2004.MOCOS05,CSRTrack}, Xsuite \cite{HB2023.TUA2I1,Xsuite}, Wake-T \cite{JPhysConfSer.1350.012056,WakeT}, and GENESIS \cite{NIMA.429.1-3.243,GENESIS}. It is also able to execute these simulations and chain multiple lattice sections together using these codes in tandem. Some examples of integrated start-to-end simulations performed using SIMBA are provided in Sec.\,\ref{sec:examples}. The translation layer is a fundamental aspect of this approach, as there is no need for the user to prepare code-specific input files; instead, based on the element and lattice definitions provided by LAURA, these objects can be translated and interacted with as Python objects, and simulations of these lattices can be executed and their outputs analyzed all within the same package.

\section{Comprehensive Format for Accelerator Lattices}\label{sec:laura}

The development of a standard accelerator format for any data structure must first address some fundamental questions. The answers to many of these questions adopted in the LAURA standard are provided in Table \ref{table:laura-standard}. 

\begin{table}
    \centering
    \begin{tabular}{|c|c|}
    \hline
    \textbf{Required definitions for lattice standard} & \textbf{LAURA definition} \\\hline\hline
    Element reference position & Middle \\\hline
    Co-ordinate system & Cartesian \\\hline
    Position definition & Design, survey and error can be included \\\hline
    Can \textit{all} accelerator types be supported? & Yes in principle, but extensive testing is needed \\\hline
    Control system information & \thead{Included in element definition} \\\hline
    Standard units & \thead{SI is used where possible;\\KnL for magnetic multipoles,\\RF phase in degrees, etc.} \\\hline
    Are beam definitions part of the lattice? & No \\\hline
    \thead{Are user stations, photon\\transport etc. part of the lattice?} & Yes, but not yet included \\\hline
    Simulation-only parameters & Defined as element attributes \\\hline
    \end{tabular}
    \caption{General questions about element and lattice definitions, and the rule adopted for LAURA.}
    \label{table:laura-standard}
\end{table}

\subsection{Element Description}\label{subsec:laura-elements}

All elements that form a lattice derive from the \texttt{baseElement} class, each instance of which must define fields including \texttt{name}, \texttt{hardware\_class}, \texttt{hardware\_type}, and \texttt{machine\_area}. The \texttt{Element} class, which derives from \texttt{baseElement}, can describe detailed information about a lattice element, such as manufacturer information, links to mechanical drawings and model files from simulation software for RF or magnet design, controls system information, and simulation-specific parameters. In this context, an \texttt{Element} can be any component that forms part of the accelerator, including elements that do not belong to a beamline, such as a klystron or laser transport system. Including these elements as part of the lattice description is useful both for storing reference information, and for understanding the connections between elements, particularly from a control system perspective. 

The \texttt{PhysicalBaseElement} derives from \texttt{Element}, and its \texttt{physical} attribute describes the position and rotation (both local and global) of the element in Cartesian space with respect to a fiducial point, which is typically defined as the location at which particles are generated. Additionally, position errors and survey measurements can be captured in the \texttt{physical} description of such an element, along with its length. Elements which are typically considered as belonging to the accelerator lattice, and as such are connected physically and contiguously with other components in a beamline along which particles travel, are all derived from the \texttt{PhysicalBaseElement} class. Depending on the \texttt{hardware\_type}, additional information is required, such as the frequency, number of cells and structure type for an RF cavity, or the bending angle for a dipole magnet; these are captured in the \texttt{cavity} and \texttt{magnetic} attributes of the \texttt{RFCavity} and \texttt{Magnet} classes, respectively. When defining an element, each of its attributes are validated so that providing invalid parameters will throw an error: a \texttt{PhysicalBaseElement} cannot have a negative length, for example. LAURA is able to describe auxiliary elements that are not typically considered as part of the lattice proper, such as low-level RF controllers or RF modulators; these elements do not derive from \texttt{PhysicalBaseElement} as they do not form part of a beamline.

\begin{figure}
    \centering
    \includegraphics[width=\linewidth]{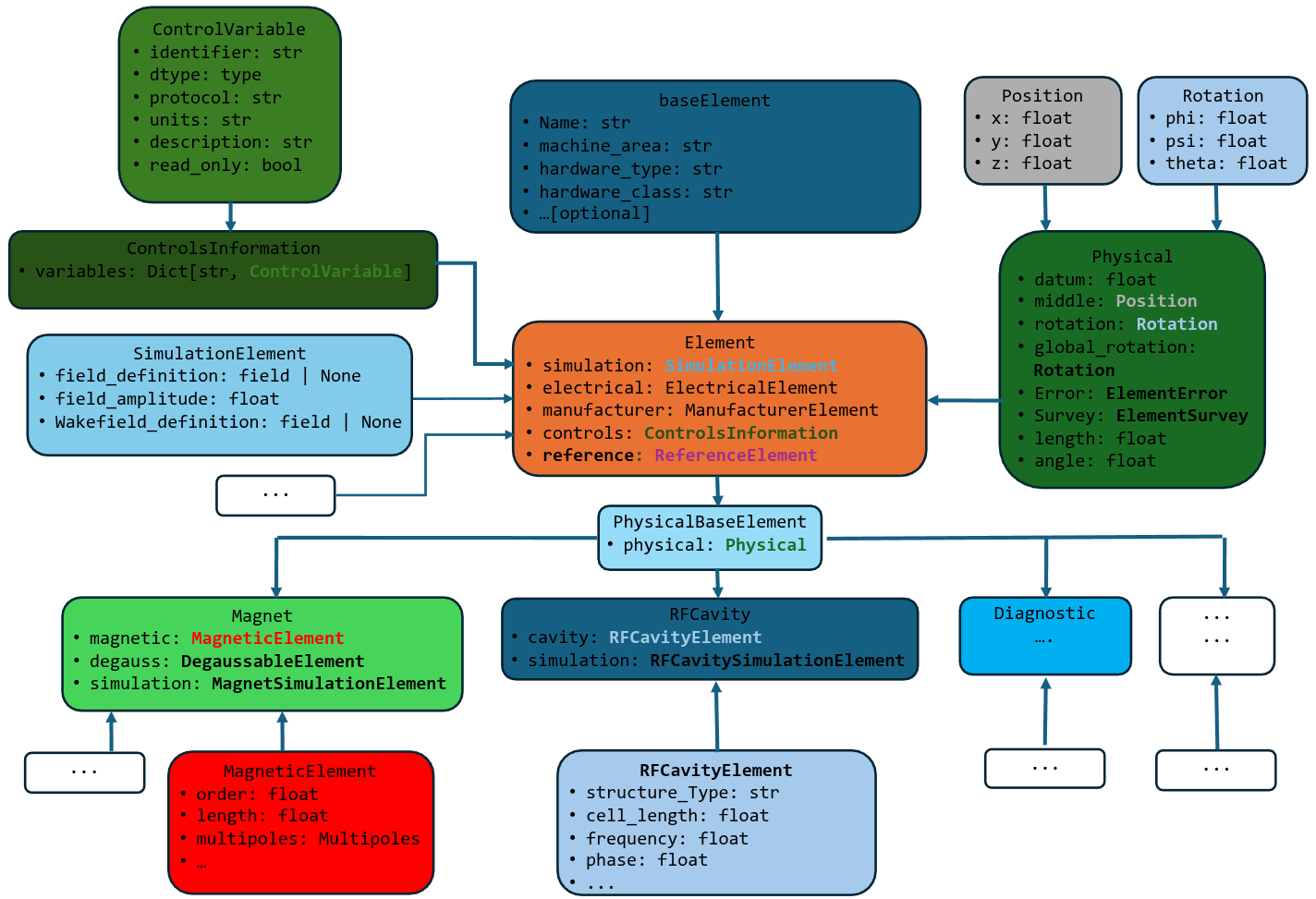}
    \caption{Flow diagram for class inheritance of LAURA elements.}
    \label{fig:element-schema}
\end{figure}

A full example of a quadrupole magnet from the CLARA accelerator \cite{PhysRevAccelBeams.23.044801} is provided in Sec.\,\ref{subsec:magnet_yaml}. This is a relatively complete description of the element, although it should be noted that it is not necessary to provide this level of detail in order to construct an element that can be used in a simulation model. The schema is sufficiently flexible to allow a full description such as this, or a minimal set of information, with only the position and magnetic strengths required. An example of how to construct a simple model of a quadrupole in Python is given in Sec.\,\ref{subsec:magnet_py}; in this case, only the positional information and quadrupole strength is required. Given that these objects can either be loaded in from a file, or created directly in Python, they can be interacted with dynamically before saving to a file or writing to a specific format for a simulation code. 

The full description in Sec.\,\ref{subsec:magnet_yaml} also shows how to link control system parameters with other attributes of the element. For example, if the control system readback provides the geometric strength of the quadrupole ($K1$), this can be written to the \texttt{magnetic.k1l} parameter of the element (the normalized magnetic strength $K1L$) via the \texttt{expression} associated with the control system variable. This schema can be generically applied as a method for converting between control system variables into units that are typically used for accelerator physics applications. 

Currently, not all conceivable accelerator lattice elements are defined within this schema, including RF quadrupoles, beam-matter interactions, and photon transport lines, although inputs from the community will be a great benefit in enabling its expansion to cover a wide user base. 

\subsection{Lattice Description}\label{subsec:laura-lattice}

With the accelerator elements now defined, it is possible to construct a lattice. In many cases, beyond small-scale machines, an accelerator can be divided up conceptually in various ways, for example depending on the beam energy in a given section. The LAURA schema defines an accelerator as consisting of `layouts', which describe the entire beam path that a particle bunch can take, and these layouts are made up of `sections'. For example, the sections in an FEL facility may include the photoinjector, the main linac, bunch compression chicanes, beam distribution switchyards, undulator sections, and photon transport beamlines. If the facility supports multiple FEL and photon transport lines, then its layouts include the common accelerating sections and the individual FEL lines, while the entire machine model encompasses all possible layouts. 

\begin{figure}
    \centering
    \includegraphics[width=0.6\linewidth]{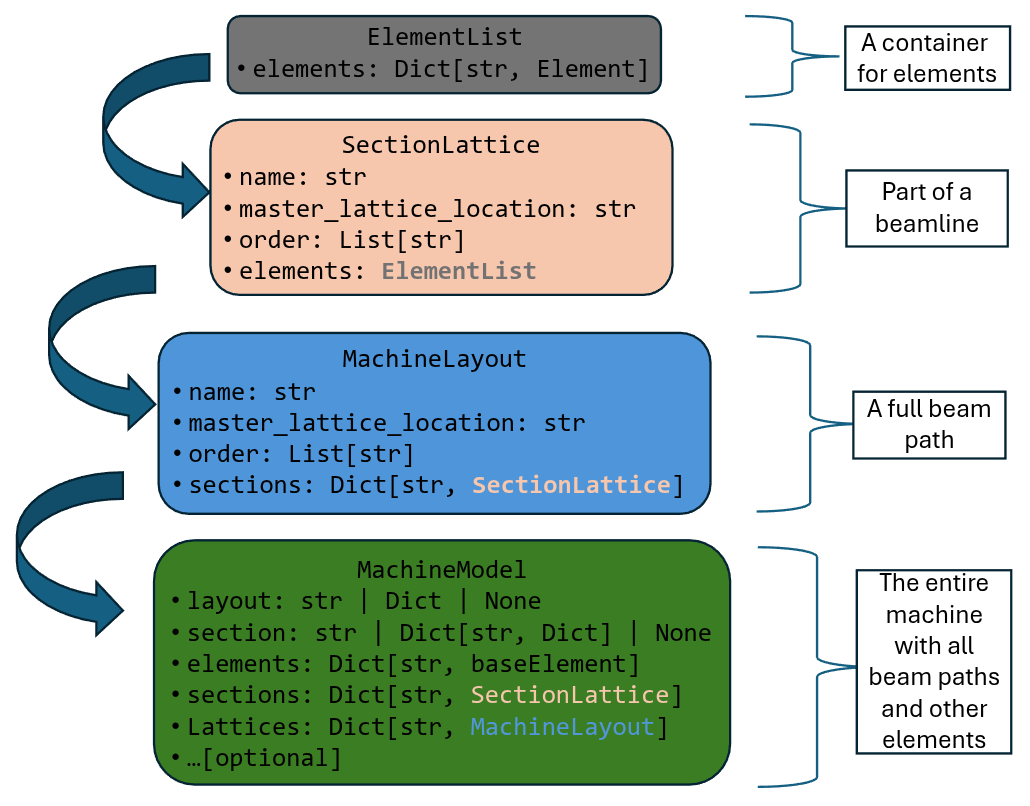}
    \caption{Flow diagram for creating sections, lattices and full machine models in LAURA.}
    \label{fig:lattice-schema}
\end{figure}

Given the descriptions of the lattice elements, these various components can be constructed. A \texttt{SectionLattice} consists of \texttt{PhysicalBaseElement} objects in a given order; a \texttt{MachineLayout} joins together contiguously connected \texttt{SectionLattice} objects; and a \texttt{MachineModel} encompasses the entire accelerator, made up of multiple \texttt{MachineLayout} elements; see Fig.\,\ref{fig:lattice-schema} for a diagram of how these components link together. In the context of start-to-end simulations, dividing the accelerator in this way enables different sections to be tracked with codes that are most suited to a given beam physics modality. 

\section{Translation Layer}\label{sec:translation}


\subsection{Writing To A Simulation Code-Specific Format}\label{subsec:lattice-export}

The comprehensive inventory of accelerator elements and beamlines described by a LAURA model can then be leveraged to write lattice files or objects that are formatted for a range of simulation codes. While this translation layer does not belong to the LAURA schema proper, it is an essential component of the start-to-end simulation framework. By deriving the lattice file from the common source of information, when examining results from various simulation codes, the user can have more confidence that they are comparing the outputs from similar inputs. 

In the translation layer, each element defined in the schema must have access to methods that write its relevant properties to a string or object for a range of simulation codes. While there is some consistency across simulation codes in terms of how the properties of elements are described, there are numerous divergences, and each code requires these descriptions to be formatted in an idiosyncratic way. Therefore, the translation layer must have access to a `dictionary' that is able to convert LAURA element attributes to a given simulation language. Once this dictionary has been created, generic methods can be employed for the majority of elements for writing to a given simulation code. In cases where a given machine element is not supported by a simulation code, warnings are typically raised to the user, and the element is passed through to the code as a drift. 

These converter functions for each accelerator element can then be called to write an entire machine section to a given format for an accelerator code. While drift sections can be defined explicitly with the LAURA schema, they are not necessary to form a lattice description, and drift sections connecting sequential elements can be created dynamically when exporting the lattice based on the distance between the end position of one element and the beginning of the next. Then, the exporter functions for a \texttt{SectionLattice} consist of iteratively calling the exporter functions to a given code for the elements in that section, including the insertion of drifts if necessary. For simulation codes which allow beamlines to be chained together, and those which support the definition of multiple layouts within a single file, exporter functions are also provided at the \texttt{MachineLayout} and \texttt{MachineModel} levels. 

Certain elements require a more detailed set of instructions in order to be translated: RF cavities, for example, may be associated with files describing its electric field and geometric wakefields for use in particle tracking, and these files must be parsed and translated. As with accelerator elements, these field files are stored in a common file format, in this case based on HDF5, which describes the electric or magnetic field of the element in up to three dimensions, along with a label for the field type. The field object can then be written into a simulation code-compliant format during the element translation process. 

\subsection{Reading From A Simulation Code-Specific Format}\label{subsec:lattice-import}

The description of accelerator elements provided by LAURA is sufficiently expansive (and extensible) that it contains enough information, in principle, to write to almost any simulation code, provided that the translation layer supports it. The reverse operation -- creating a LAURA lattice based on a lattice file for an existing simulation code -- is more complicated. Certain simulation codes provide to the user a great deal of flexibility in how elements are defined, including aliases for parameter names, and the definition of element attributes as expressions rather than numerical values. A generic solution for this half of the translation layer is therefore much more difficult to achieve. 

For many simulation codes, the user is able to output the lattice in a standardized format as a result of tracking, which reduces the scope of possible items that need to be translated. Therefore, the lattice interpretation layer typically makes use of these output files from various simulation codes, and in general the conversion to LAURA relies on reversing the names in the conversion dictionary used for exporting lattices to code-specific formats. It may be the case that not all elements provided by a given code are defined in this dictionary, in which case the user is typically warned. Given that elements are defined in Cartesian co-ordinates in the LAURA standard, it is helpful if element floor co-ordinates can be produced as a result of tracking, although tools are provided for translating from Frenet-Serret to Cartesian. 

Finally, as mentioned above, electromagnetic fields associated with an element should be included in a standard format so that it can be translated into a variety of codes. However, when reading in such an object from a specific code, translation of that field object into a generic format is required. While some helper functions are provided to interpret various field files, it is recommended that, during interpretation of a given lattice file, the user ensures that the conversion of these fields is done correctly, including the definition of the field type and the field strength units.

\section{Simulation Interface}\label{sec:simba}

The development of a translation library for accelerator elements and lattices provides a fundamental step for establishing a generic start-to-end simulation framework, although a further set of commands are needed to enable seamless handshaking between codes during tracking. Additional necessary concerns include the handling and generation of particle beam distributions, the preparation of input files and objects, and the execution and analysis of simulations. All of this functionality can be handled by a central orchestrator -- a start-to-end simulation framework, SIMBA \cite{SIMBA}. 

\begin{figure}
    \centering
    \includegraphics[width=\linewidth]{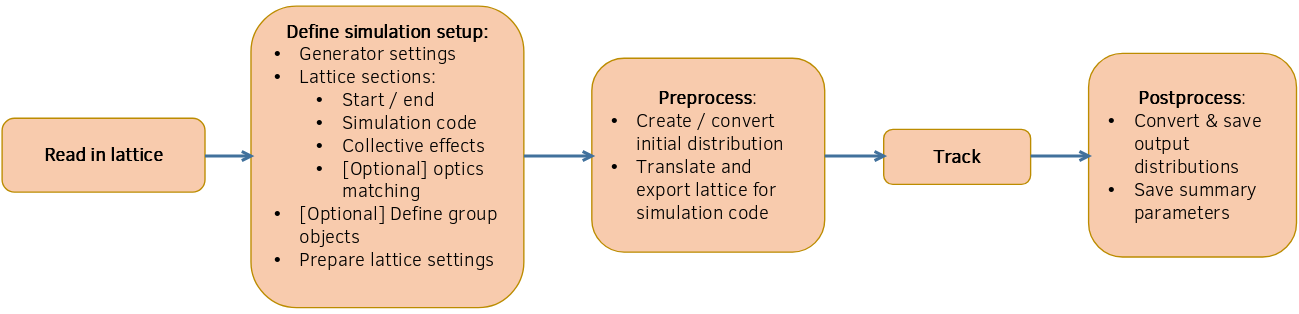}
    \caption{Flow diagram demonstrating how to prepare, execute and analyze a simulation in SIMBA.}
    \label{fig:simba-flowchart}
\end{figure}

\subsection{Orchestration}\label{subsec:framework}

SIMBA is a Python package that is able to read in lattice files from LAURA and create separate lattice lines dynamically based on an input file. From this lattice, the user can define which simulation codes are to be used for given sections of the accelerator, and track through them. Before tracking, lattice elements can be adjusted dynamically, rather than configuring settings in input files. Both the lattice elements and full lattice objects are accessible as attributes of the SIMBA instance, and the elements can be controlled via their LAURA names and attributes. Additionally, simulation settings can be configured, such as the computation of certain collective effects that are relevant for that section, the setting of initial beam Twiss parameters, or matching targets at specific locations; other options, such as for the preparation of a FEL or plasma-based simulation, can also be configured. Elements can also be grouped together in order to perform co-ordinated operations on them as if they were a single element, such as the setting of bending angles and entrance/exit angles for multiple dipoles in a magnetic chicane; future implementations could include setting the optics for a multi-bend achromat section.

The entire lattice, or sections within it, can then be tracked sequentially; see Fig.\,\ref{fig:simba-flowchart} for a flow diagram of how the information provided by the LAURA lattice and simulation definition file is used to prepare and execute a simulation. An initial beam distribution is either loaded in or generated, and both this data and the lattice section are converted to the correct format for the simulation code defined. All manipulations of the lattice or particle distribution can be performed as part of a script before the simulation files are constructed using the methods described in Sec.\,\ref{subsec:lattice-export}. The simulation is then executed, and the output data are saved ready for the next step of the simulation. For computationally-intensive lattice sections, or for full simulation runs, the user can specify the location of a computer cluster on which to run the simulation. In this case, the prepared files are all sent to the cluster, executed remotely, and sent back to the main SIMBA instance for the next tracking step. Summary information for a section includes the Twiss parameters along the line, and links to the beam distributions that are produced during tracking. After the full simulation is complete, the lattice files, along with the settings provided to the simulation, are stored along with the beam distributions. 

It should also be noted that the tracking codes currently supported provide a wide range of possible simulation configurations, some of which are incompatible with the other supported codes. Given both that it would be unfeasible at this stage to enable all possible configurations of all simulation codes within the package, and that it could greatly complicate the simulation input file format, only a reduced set of generic options are currently provided to the user of SIMBA, such as the enabling and toggling of collective effects, tracking step sizes, and the setting of beam matching parameters. If more detailed code-specific simulation options are required, the user can load in a particular lattice section and produce input files, and then modify them offline; alternatively, for Python-based simulation codes, these options can be configured dynamically. Future developments will take into consideration which of these options to expose in the top-level configuration file.

\subsection{Phase Space Distributions}\label{subsec:beams}

The particle beam distribution to be used during tracking is available as a global parameter in SIMBA. In addition to the 6D macroparticle distribution (based on SI units), information such as the particle mass and species, filename, and the code from which the distribution was produced, are stored. Analysis functions are also available for computing slice parameters, centroids, Twiss, emittance, and other statistical properties of the beam. One widely adopted standard file format for beam distributions is \texttt{OpenPMD-beamphysics} \cite{OpenPMD-beamphysics}, a widely-used standard based on HDF5. A number of simulation codes employ this standard as their I/O format, while for other codes, conversion is required. This is done in SIMBA during pre- and post-processing of each tracking step. The phase space representation used internally for SIMBA expands on the standard \texttt{OpenPMD-beamphysics} format while maintaining its core definitions: in addition to the 6D beam distribution and macroparticle charge and species, the longitudinal s-position and reference particle are also stored and transferred between codes to ensure consistency. Most simulation codes include diagnostic screens or markers as elements at which the phase space distribution is saved to a file; the post-processing routine for each section during tracking converts all of these distributions into the standard file format. 

\subsection{Beam Generation}\label{subsec:generator}

Phase space distributions can be created independently of particle tracking and injected into the simulation, or the user can specify \texttt{generator} parameters in their settings file, which will prepare a distribution before tracking. Two options are currently available for the generation of initial particle distributions. First, the user can generate semi-arbitrary phase spaces based on Gaussian or flat-top distributions in each plane, and add longitudinal energy chirps; electrons, positrons, protons and hydrogen ions are currently supported. Second, electron distributions can be produced from a photocathode and generated using either ASTRA \cite{ASTRA} or GPT \cite{GPT}. In this case, additional parameters can be provided, including the cathode parameters such as its work function and Fermi energy, and the laser transverse and longitudinal profile. These generated distributions are then saved into the standard format before being passed to a simulation code. 

\subsection{Optimization}\label{subsec:optimisation}

As with beam distributions, SIMBA leverages existing tools that are widely used by the accelerator community for the optimization of beam parameters; in this case, \texttt{Xopt} \cite{Xopt,IPAC2023.THPL164} is available as a generic optimization package. This is a flexible, high-level package that supports a range of algorithms, and provides a comprehensive analysis toolkit. The wrapper for SIMBA reads in the constraints and variables from the \texttt{Xopt} input file, prepares the simulation and analyzes the output. The benefit of using SIMBA for these tasks is that beam optimization can be done at an arbitrary scale: for a large facility such as an X-ray FEL, for example, the quality of the radiation produced depends on a wide range of parameters, starting with the initial beam distribution generated at the photocathode. A full optimization of the FEL must take all of these parameters into account, and SIMBA allows the user to focus on the physics in the simulation rather than preparing input files for these various machine sections during optimization. 

\section{Examples}\label{sec:examples}

\begin{figure}
    \centering
    \includegraphics[width=\linewidth]{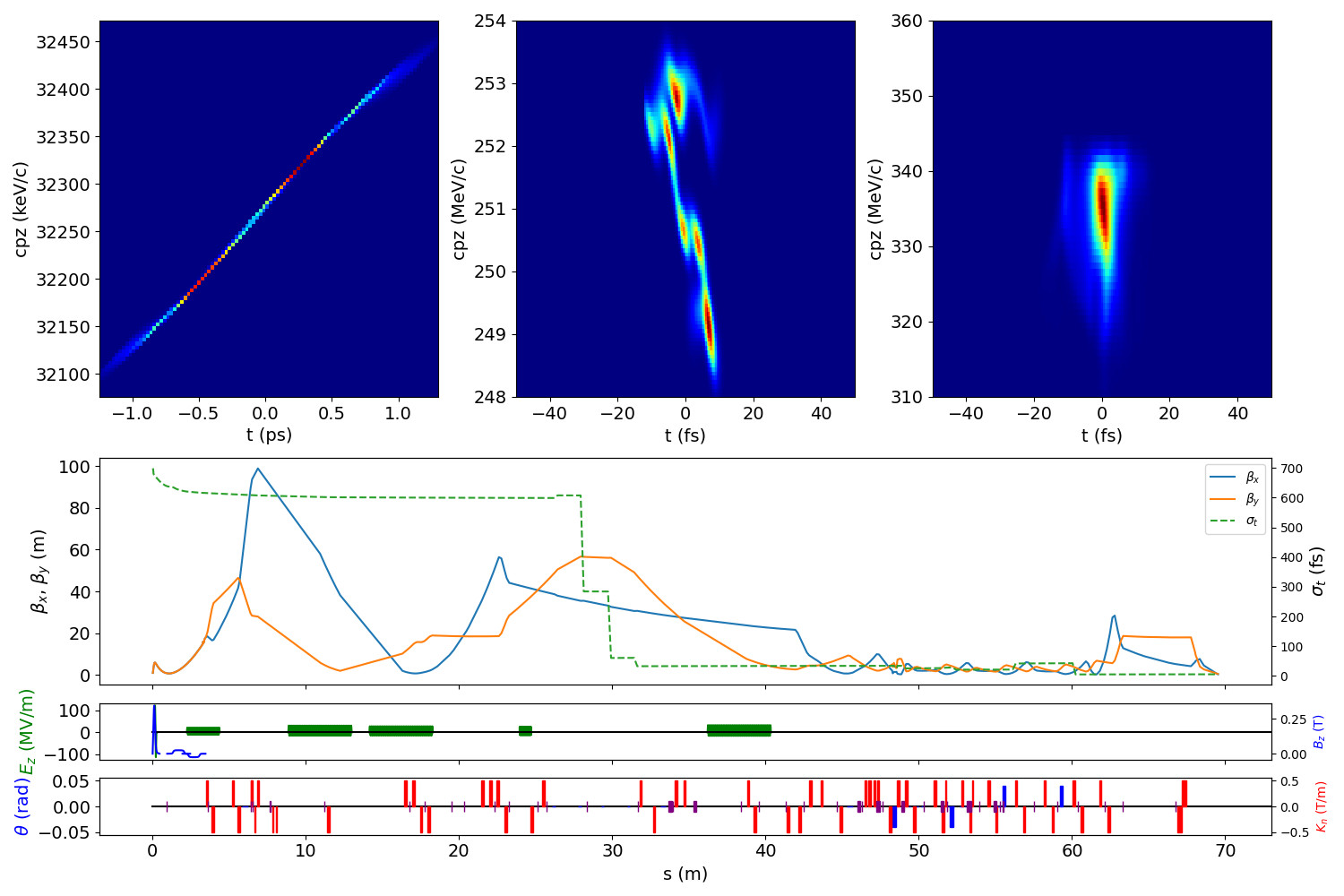}
    \caption{Full start-to-end simulation of the CLARA accelerator using SIMBA. The beam is generated at the photocathode and tracked to the end of the first linac using ASTRA (top left figure). The beam is then matched transversely using OCELOT and tracked through the remainder of the Linac and the FEBE arc using ELEGANT (top middle figure). Finally, the beam is sent through a plasma stage using Wake-T (top right figure). The beta functions and bunch length for the entire machine are shown in the bottom plot.}
    \label{fig:clara_s2e}
\end{figure}

A full simulation of the CLARA accelerator and the FEBE arc line \cite{PhysRevAccelBeams.23.044801,PhysRevAccelBeams.27.041602}, performed using SIMBA and based on a LAURA machine model, is shown in Fig.\,\ref{fig:clara_s2e}. A $5$\,\si{\pico\coulomb} beam was generated at the photocathode and tracked through the RF gun and first linac using ASTRA, then transverse matching and tracking through the linac was performed using OCELOT. The beam is overcompressed in the variable bunch compressor at around $25$\,\si{\metre} in order to reach a highly compressed beam at the exit of the arc line beginning at $48$\,\si{\metre}. The matched beam was then tracked through to the FEBE user station with ELEGANT, and Wake-T was used to simulate laser-plasma acceleration using the FEBE terawatt laser. As described above, all handling of the simulation input file generation, machine setup, tracking, and beam pre- and post-processing was done using the in-built functionality of SIMBA. 

\begin{figure}
    \centering
    \includegraphics[width=\linewidth]{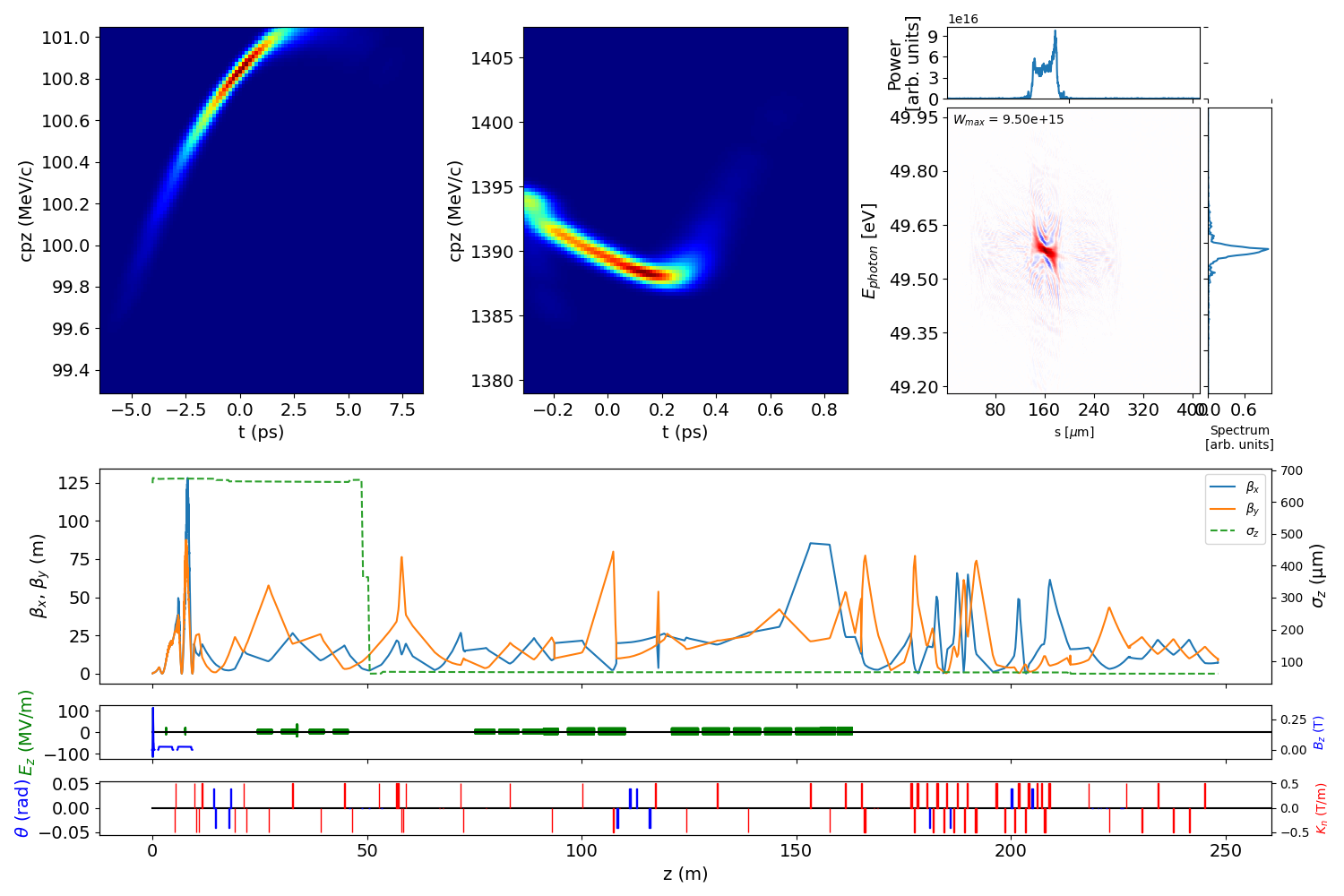}
    \caption{Full start-to-end simulation of the FERMI accelerator and FEL using SIMBA. The beam is generated at the photocathode and tracked through the first two linac using ASTRA (top left figure). The beam is then tracked through the linac, bunch compressor and spreader using OCELOT (top middle figure). Finally, GENESIS is used to simulate the external seeding and generation of FEL radiation (top right figure). The beta functions and bunch length for the entire machine are shown in the bottom plot.}
    \label{fig:fermi_s2e}
\end{figure}

Figure\,\ref{fig:fermi_s2e} shows an analogous set of results for the FERMI FEL \cite{NatPhoton.6.699}, for which the LAURA instance was created based on an ELEGANT lattice file: the beam generation and linac tracking was again performed with ASTRA and ELEGANT, and the FEL simulation was conducted with GENESIS. Given that FERMI is a seeded FEL, both the parameters for the seed laser and the post-modulator delay chicane must also be specified with LAURA. Despite the fact that the inputs required to generate a GENESIS simulation differ from those used for standard beam dynamics codes, SIMBA offers many of these features to the user. 

\section{Conclusions}\label{sec:conclusions}

This article has described both a standard format for particle accelerator lattices, and a framework for generating start-to-end simulations based on this format. The objects defined in this format cover a wide range of description formats for each element, including physical data (with positional errors), simulation code-specific settings, control system parameters, and electrical, manufacturer and reference information, in addition to element-specific information pertaining to particular types. While the current set of objects described in the LAURA format does not cover all possible elements that feature in all different types of accelerators, many of the most common types are defined in the standard, providing a strong base for further work. 

One benefit of defining accelerator elements and lattices in a generic format such as this is that it does not rely on a lattice file from a particular simulation code, or any other mode of description, as a ground source of truth. The in-built functionality of LAURA to generate simulation code input files is leveraged by the SIMBA package, which provides an abstraction layer to simulations and can therefore be used to orchestrate start-to-end simulations. Given that both element-specific parameters such as magnet strengths can be modified dynamically with a LAURA instance, and that simulation-specific parameters can also be set, SIMBA allows a flexible interface to a wide configuration space for start-to-end simulations. External packages for describing beams, and for optimizing machines, are also employed by SIMBA, meaning that it can be useful for many accelerator facilities. As with the LAURA package, not every accelerator code is supported by SIMBA, and the user only has access to a restricted range of possible settings for all codes. These can be expanded based on demand; in its current state SIMBA is sufficiently flexible to be able to simulate a range of beam dynamics modalities and machines, although it has not yet been stress-tested on multi-pass accelerators, for example. Alternatively, the PALS project, while in an earlier stage of development with respect to LAURA, plans to cover a more expansive range of particle accelerator types from the outset. In the future, provided that the translation layer to simulation code formats is sufficiently robust, PALS-style lattices could instead be loaded into SIMBA, with the same beam input/output and tracking functionality maintained.

In the meantime, given that control system parameters are fully described for LAURA elements, a clear next step is integrating start-to-end simulations with virtual control systems. The CATAP package \cite{arXiv.2509.19794} and other control system middle layer projects such as pyAML \cite{pyAML} could benefit from a lattice format that is able to communicate both with physics- and simulation-based parameters and control system variables. Work is currently under development to use a LAURA lattice both to generate a full virtual control system, and to connect a simulation model to this via SIMBA in order to provide a path towards a digital twin architecture that can be used by any accelerator facility. Furthermore, automated workflows and agentic artifical intelligence applications can leverage this comprehensive description of control system variables, and how they link to the machine physics model, to provide a generic core ontology for accelerators \cite{arXiv.2512.00868}.

\bibliographystyle{unsrtnat}
\bibliography{references}

\pagebreak
\appendix\label{appendix}

\section{Example Lattice Element} 
\label{sec:example_element_file}

\subsection{CLA-S02-MAG-QUAD-01.yaml}
\label{subsec:magnet_yaml}

\begin{lstlisting}[language=yaml]
# In CLARA/YAML/Magnet/Quadrupole/CLA-S02-MAG-QUAD-01.yaml
alias: null
controls:
  variables:
    readback:
      description: Magnetic strength of CLA-S02-MAG-QUAD-01
      dtype: float
      identifier: CLA-S02-MAG-QUAD-01:RBV
      protocol: CA
      read_only: true
      units: N/A
      target: magnetic.k1l
      expression:
        op: mul
        args:
          - readback
          - magnetic.length
    setpoint:
      <...>
    <...>
degauss:
  steps: 11
  tolerance: 0.5
  values: 
    - 52.89
    - -52.89
    - 31.8
    <...>
electrical:
  maxI: 53
  minI: -53.0
  read_tolerance: 0.1
hardware_class: Magnet
hardware_type: Quadrupole
machine_area: S02
magnetic:
  field_integral_coefficients:
    coefficients: 
      - -2.23133410405682e-10
      - 4.5196171252132e-08
      <...>
  length: 0.12868478212775
  linear_saturation_coefficients:
    I0: 942.4766071889139
    I_max: 15.575435287034225
    <...>
  multipoles:
    K1L:
      normal: 1.79019
      order: 1
      radius: 0.0
      skew: 0.0
  order: 1
  random_multipoles: {}
  settle_time: 45.0
  skew: false
  systematic_multipoles: {}
manufacturer:
  hardware_class: Quadrupole
  manufacturer: Quadrupole Type 1
  serial_number: '13248'
name: CLA-S02-MAG-QUAD-01
physical:
  datum: [0.0, 0.0, 3.591]
  error:
    position: [0.0, 0.0, 0.0]
    rotation: [0.0, 0.0, 0.0]
  global_rotation: [0.0, 0.0, 0.0]
  length: 0.1007
  middle: [0.0, 0.0, 3.527]
  rotation: [0.0, 0.0, 0.0]
  survey:
    position: [0.0, 0.0, 0.0]
    rotation: [0.0, 0.0, 0.0]
subelement: false
virtual_name: ''
\end{lstlisting}

\pagebreak

\subsection{QUAD01.py}
\label{subsec:magnet_py}

\begin{lstlisting}[language=Python]
from laura.models.element import Quadrupole

quad_dict = {
    "name": "QUAD1",
    "machine_area": "FODO",
    "magnetic": {
        "order": 1,
        "length": 0.1,
        "kl": 1.0,
    },
    "physical": {
        "middle": [0, 0, 0.1],
        "length": 0.1,
    },
}

q1 = Quadrupole(**quad_dict)
\end{lstlisting}

\pagebreak

\end{document}